\documentclass[showpacs,floatfix,aps]{revtex4}
\usepackage{amssymb,amsmath}
\usepackage[dvips]{graphics,color}
\usepackage{epsfig}
\usepackage{float}
\usepackage[T1]{fontenc}
\usepackage{lscape}
\usepackage{cleveref}

\usepackage{lscape}
\newcommand{\bea}{\begin{array}}
\newcommand{\ear}{\end{array}}

\newcommand{\bege}{\begin{equation}}
\newcommand{\enge}{\end{equation}}

\newcommand{\beq}{\begin{eqnarray}}\newcommand{\benu}{\begin{enumerate}}\newcommand{\enu}{\end{enumerate}}
\newcommand{\eeq}{\end{eqnarray}}

\newcommand{\noi}{\noindent}

\begin{document}

\title{Luminosity of ultra high energy cosmic rays and bounds on magnetic luminosity of radio-loud AGNs}

\author{C. H. Coimbra-Ara\'ujo}
\email{carlos.coimbra@ufpr.br}
\author{R. C. Anjos}
\email{ritacassia@ufpr.br}


\affiliation{Departamento de Engenharias e Ci\^encias Exatas,
  Universidade Federal do Paran\'a (UFPR),\\ 
Pioneiro, 2153, 85950-000 Palotina, PR, Brazil.}


\pacs{13.85.Tp, 98.70.Sa, 95.30.Qd, 98.62.Js}

\begin{abstract}

We investigate the production of magnetic flux from rotating black
holes in active galactic nuclei (AGNs) and compare it with the upper limit of 
ultra high energy cosmic ray (UHECR) luminosities, calculated from observed integral flux of GeV-TeV gamma rays 
for nine UHECR AGN sources.
We find that, for the expected range of black hole rotations ($0.44<a<0.80$), 
the corresponding bounds of theoretical magnetic luminosities from AGNs coincides 
with the calculated UHECR luminosity. We argue that such result possibly can  
contribute to constrain AGN magnetic and dynamic properties as phenomenological tools  
to explain the requisite conditions to proper accelerate the highest energy cosmic rays.

\end{abstract}

\maketitle

The origin of the highest energy cosmic rays, or ultra high energy cosmic rays (UHECRs), with energies $E > 10^{18}$ eV, 
in spite of huge observational/experimental endeavors, 
represents one of the greatest puzzles of modern astrophysics \cite{olinto10,lemoine12}. 
The possibility of accelerating particles up to such extreme energies is addressed by the well-known Hillas source plot \cite{hillas84}.
There, a table of possible UHECR sources is presented by bringing the simple yet efficient remark that particles, during acceleration, 
are confined in the source on a Larmor timescale (for more details see, e.g., \cite{nagano1,nagano2,bluemer,beatty,letessier,olinto10,lemoine12,watson13}).

The prominent extragalactic candidates for accelerating particles (mainly protons) to the highest energies are active galactic nuclei (AGNs) \cite{henri}, 
the most powerful radiogalaxies \cite{takahara,rachen}, and also gamma ray bursts, fast spinning newborn
pulsars, interacting galaxies, large-scale structure formation shocks and some other objects \cite{olinto10}.

On the other hand, the reconstruction of cosmic ray luminosities, from Earth laboratory observations, possibly can shed some light on radiative bounds of UHECR potential sources. 
For example, in \cite{supa,vitor} it is shown that using the methods of UHECR propagation from the source 
to Earth and the measured upper limit on the integral flux of GeV-TeV gamma-rays it is possible to infer the upper limits of the proton and total UHECR 
(iron) luminosity. This comes from the fact that gamma-rays can be produced as a result of the cosmic ray propagation and contribute to the total flux measured
from the source.

In the present work, some of the potential AGN sources will be investigated as UHECR sources (see Table \ref{tab:1}).
It will be calculated upper limits of UHECRs luminosities to be compared to the theoretical magnetic/jet luminosity of those AGNs. 
To reconstruct, from experiments, the possible UHECR luminosities here it will be used the method described in \cite{supa} 
as a prolific way to calculate upper limits on UHECR luminosities. 

In first place, space and ground instruments, as FERMI-LAT \cite{fermi}, VERITAS \cite{veritas}, H.E.S.S. \cite{hess} and MAGIC \cite{magic}
provide upper limits on the GeV-TeV gamma-ray integral flux and the
method \cite{supa} connects those measured upper limits with the source UHECR cosmic ray luminosity ($L_{CR}^{UL}$) by 
\begin{equation}
L_{CR}^{UL} = \frac{4\pi D^{2}_s(1+z_s)\langle E \rangle_{0}}{ \ K_{\gamma} {\mathop{\displaystyle
\int_{E_{th}}^{\infty} dE_\gamma\ P_{\gamma}(E_{\gamma})}}}\ I_{\gamma}^{UL}(> E^{th}_{\gamma}),
\label{eq:CRUL}
\end{equation}
where $I_{\gamma}^{UL}(> E^{th}_{\gamma})$ is the upper limit on the integral gamma-ray flux for a given confidence level and energy
threshold, $K_{\gamma}$ is the number of gamma rays generated from the
cosmic ray particles, $P_{\gamma}(E_{\gamma})$ is the energy
distribution of the gamma-rays arriving on Earth, $E_{\gamma}$ is the
energy of gamma-rays, $\langle E \rangle_{0}$ is the mean energy, $D_s$
is the comoving distance and $z_s$ is the redshift of the source. This
method allows one to calculate upper limits on the proton and total
luminosities for energies above $10^{18}$ eV. Also, it illustrates techniques to study the origin of UHECRs
from multi-messenger GeV-TeV gamma-rays and it has been used to calculate
at least upper limits for thirty sources (AGNs), with redshift smaller
than 0.048 and UHECR spectra measured by the Pierre Auger \cite{pierre} and Telescope
Array \cite{ta} (TA) observatories. In fact, as described by \cite{supa,vitor}, the UHECR upper limit 
luminosity is obtained from the integral of the gamma-ray flux from the observed spectrum of UHECRs. 
Propagation models and the measured upper limit of gamma-ray flux of a source amongst propagation 
models are the fundamental aspects to be considered to perform such calculation. Indeed, this same method will allow, 
in the future, from CTA Observatory \cite{cta}, a range of new UHECR luminosity upper limits. 

In second place, it is well-known that radiative mechanisms within AGNs are chiefly powered by their central supermassive black holes (SMBHs) and the companion accretion disk. 
The rotation of the system creates twisted magnetic fields that drive jets of relativistic particles. Here it will be considered a standard accretion 
mechanism (Bondi accretion model $\dot{M} = \pi \lambda c_{S} \rho_B r_B$ \cite{bondi}, 
where $\dot{M}$ is the accretion rate, $\lambda$ takes the value of 0.25 for an adiabatic index 5/3, $c_S$ is the sound speed in the medium, 
$r_B$ is the Bondi accretion radius and $\rho_B$ is the gas density at that radius) which produces, by friction and other radiative processes, enormous bolometric luminosities. The injection 
of only a moderate fraction of this bolometric luminosity would suffice to reproduce the observed cosmic ray flux above $10^{19}$ eV. 
Nevertheless, it appears that cosmic ray flux from AGNs probably comes from jet luminosities. Therefore, the magnetic field of the structure is 
the main responsible to produce such perpendicular jet outflows. In this case, they do not all offer as appetizing physical conditions with 
respect to particle acceleration as, e.g., gamma ray bursts, since 
considerable outflow luminosities, i.e., magnetic luminosities $L_B$ are actually required to accelerate protons to the highest energies observed. 
This limits the potential number of AGNs as cosmic ray sources in the nearby universe (unless the highest energy cosmic rays are heavy nuclei) \cite{lemoine12}. 
In the present contribution, we describe the mechanism via powering jets from Blandford-Znajek mechanism \cite{marek}. 
This approach is based on flux accumulation that leads to the formation of a
magnetically chocked accretion flow and strong flux of high energy particle jets and it assumes that any geometrically thick or hot inner region of an accretion flow 
can drive magnetic field fluctuations to produce jets. This technique has been proposed to explain the most luminous and radio-loud AGNs, 
and consequently could also explain mechanisms behind the production of UHECRs.

In what follows it is derived a description for the production of UHECR luminosities based on
the possible relation between magnetic flux accumulation and jet production
efficiency. Similar schemes have been proposed to describe UHECR luminosity
contribution and sources of cosmic rays from black hole accretion mechanisms
\cite{ioana, biermann}. The significant feature of this model is that the dominant factor in the magnetic luminosity
is the powerful jets to therefore determinate the radio loudness of the AGN.	
In first place, considering a system with a rotating central BH, the necessary condition to use Blandford-Znajek mechanism is that
$\Phi_d > \Phi_{BH,\hbox{max}}(\dot{M})$, i. e., the net poloidal magnetic flux $\Phi_d$ trapped in the disk
is larger than the maximum that can be confined on the BH caused by pressure of the accreting plasma. Satisfied this condition, 
the rate of energy extraction from the rotating BH via the Blandford-Znajek mechanism yields the magnetic luminosity

\begin{eqnarray}\label{lb}
L_{B} &\simeq& 4\times 10^{-3}\Phi_{BH,\hbox{max}}^{2}(\dot{M})\frac{\Omega_{BH}^{2}}{c}f_a(\Omega_{BH})\nonumber\\
&=& 10(\phi/50)x_{a}^{2}f_{a}(x_{a})\dot{M}c^{2},
\end{eqnarray}\noindent 
where $\Omega_{BH}$ is the angular velocity of the black hole and
\begin{equation}
x_a=r_{g}\frac{\Omega_{BH}}{c}=[2(1+\sqrt{1-a^{2}})]^{-1}a,
\end{equation}
with
\begin{equation}
f_{a}(x_a)\simeq 1 + 1.4x_{a}^{2}-9.2x_{a}^{4},
\end{equation}
where $a$ is the dimensionless angular momentum parameter ($a=J/Mc$, with $J$ the BH angular momentum), $\phi$ 
is a dimensionless factor which, according to numerical simulations (see, e.g., \cite{mckinney}), is typically of order 50, and 
$r_g$ is the gravitational radius $r_g=GM/c^2$.

\begin{figure}[h]
  \centering
  \includegraphics[width=9cm]{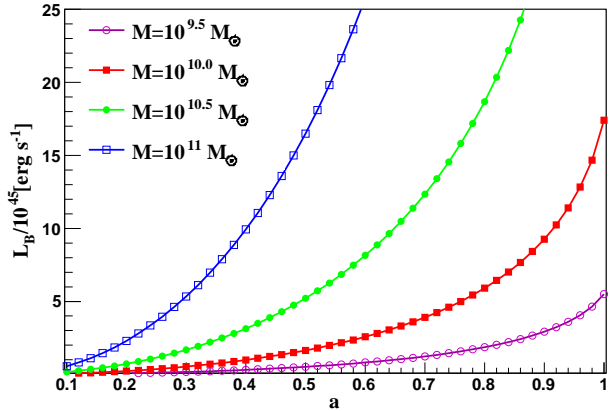}\\
\caption{$L_{B}$ as a function of the mass and parameter
  $a$. For all plots it was considered $\dot{M}$ as the Bondi accretion rate \cite{bondi}. For comparison, the Eddington luminosities $L_{\mathrm{Edd}}$
  for each considered AGN mass are: $L_{\mathrm{Edd}}(M=10^{9.5}M_\odot)=1.26\times10^{47.5}$ erg s$^{-1}$, $L_{\mathrm{Edd}}(M=10^{10.0}M_\odot)=1.26\times10^{48}$ erg s$^{-1}$,
  $L_{\mathrm{Edd}}(M=10^{10.5}M_\odot)=1.26\times10^{48.5}$ erg s$^{-1}$, $L_{\mathrm{Edd}}(M=10^{11.0}M_\odot)=1.26\times10^{49}$ erg s$^{-1}$.}
\label{fig:luminosity1}
\end{figure}


In Fig. \ref{fig:luminosity1} it is displayed the effect of black hole spinning on magnetic luminosity with
the variation of mass. The greater the AGN BH spinning, the greater the 
AGN magnetic loudness. A similar result is observed with the increasing of the black hole mass. In this approach, $M$ and $a$ are therefore the main 
parameters to affect energetic jets and the consequent energetic cosmic ray production at AGN expected sources. For comparison one can calculate 
the upper limit of bolometric luminosity for each source, i.e., the Eddington luminosity $L_{\mathrm{Edd}}=\frac{4\pi G M m_p c}{\sigma_T}= 1.26\times10^{45}\frac{M}{10^7M_\odot}$ 
(where $M$ is the mass of the AGN, $m_p$ is the hydrogen atom mass, $c$ is the light speed and $\sigma_T$ is the Stephen-Boltzmann constant) to see that, as awaited, for each 
source $L_B < L_{\mathrm{Edd}}$.


As AGN jets can be the main extragalactic sources of UHECRs
\cite{henri,biermann1}, one can write the UHECR luminosity $L_{CR}^{Theory}$ as a
fraction $\eta$ of the magnetic luminosity: 
\begin{equation}\label{eq:lumicosmic}
L_{CR}^{Theory} = \eta L_{B}.
\end{equation} 
\noi For example, bounds on the fraction $\eta_{pr}$ of $L_B$ to be converted in relativistic protons, as a function of BH spin $a$, come 
from eqs. (\ref{eq:CRUL}) and (\ref{eq:lumicosmic}) as
\begin{equation}\label{eq:etapr}
\eta(a)_{pr} = \frac{L_{pr}^{UL}}{10(\phi/50)x_{a}^{2}f_{a}(x_{a})\dot{M}c^{2}}. 
\end{equation}

\begin{figure}[h]
  \centering
  \includegraphics[width=9cm]{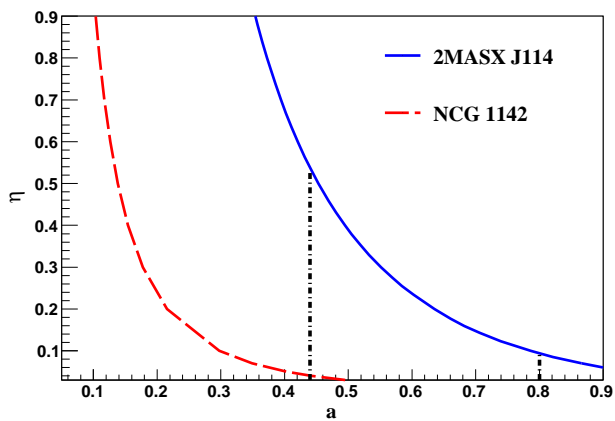}\\
\caption{$\eta_{pr}$ as a function of the parameter
  $a$. For all plots it was considered $\dot{M}$ as the Bondi accretion rate \cite{bondi}.}
\label{fig:n_versus_a}
\end{figure}

For the spin range $0.44 < a < 0.8$ where most black holes
are expected to lie, the fraction $\eta_{pr}$ varies from as $\sim 5\%$ to $40\%$ for $a=0.45$ and from $2\%$ to $10\%$ for $a=0.8$, see Fig. \ref{fig:n_versus_a}.
An important remark is that Fig. \ref{fig:n_versus_a} reflects the outter bounds of all nine curves $\eta(a)_{pr}$ produced from the nine sources of Table \ref{tab:1}. 
The sources 2MASX J1145 and NGC 5995 are taken as the critical limits for AGNs in the universe up to redshifts $z_s < 0.048$ and they 
plot the threshold curves for eq. (\ref{eq:etapr}). See, for instance, that the upper limit $L^{UL}_{pr}$ for these two sources is greater
than the theoretical calculated range (for a spin $a \sim 0.7$). In this case, it is expected that their masses produce low levels of magnetic luminosity. 
To explain such unexpected great upper limits, one has to admit, e.g., that possibly those AGNs have a critical spinning black hole, with 
$a\rightarrow 1$, since the greater the spin, the greater the luminosity $L_B$.

Table \ref{tab:1} shows the range $L_{CR\hbox{,min}}^{Theory} - L_{CR\hbox{,max}}^{Theory} = \eta_{pr,\hbox{min}}L_B - \eta_{pr,\hbox{max}}L_B$ for nine 
AGN sources, assuming a fixed spin of $a=0.45$ and $a=0.7$. The column $L_{pr}^{UL}$ is the calculated upper limit of the proton luminosity for each source from 
eq. (\ref{eq:CRUL}), and it has dependence only with observational [GeV-TeV gamma rays + cosmic rays] constraints \cite{vitor,supa},
i.e., it has no dependence with BH mass or BH spin of the AGN.   
Comparing both results, it is possible to see that NGC 1142 fails to produce a proton upper luminosity that remains in the $L_{CR}^{Theory}$ 
calculated range. This is explained by the fact that the observed
NGC 1142 mass must drive a great magnetic luminosity $L_B$, since the greater the 
mass, the greater the luminosity $L_B$. To explain the low cosmic ray
luminosity from NGC 1142 one can assume that the AGN black hole possibly has a small 
spin, since the smaller the spin, the smaller the luminosity $L_B$ and consequently the smaller the cosmic ray luminosity. 

The present work compared ultra high energetic cosmic ray luminosities with a theoretical luminosity derived from intrisic AGN properties, 
as its mass and central black hole spin. An important remark is that luminosity calculation from method (\ref{eq:CRUL}) does not require any AGN property, 
since they come unically from observed integral flux of GeV-TeV gamma-rays of UHECR AGN sources.
In this aspect, both for proton and iron luminosities it is possible to find phenomenological bounds on the conversion fraction of magnetic luminosities into 
energetic particles. Such results logically lead to the question whether AGNs are or not the sources of the high energy cosmic rays. 
In general, the Fermi acceleration processes are evoqued to explain if AGNs could indeed reach $\sim 10^{19.5}$ eV. In parallel, 
for protons, only the most powerful flat spectrum radio quasars, which are considered to be the jet-on analogs of FR II radio galaxies with relativistic jets,
show a magnetic luminosity in excess of $10^{45}$ ergs s$^{-1}$ that can power sufficient energetic jets. Those are the objects studied here. 
In complement, BL Lac objects or TeV blazars, thought to be the analogs of FR I radio galaxies (such as Cen A), typically exhibit magnetic luminosities $L_B$ of the order of $10^{44}$ ergs s$^{-1}$ or less. 
In such cases, only Fermi processes and shocks in the jets and the hot spots of the most powerful FRII 
radio-galaxies may nevertheless offer the requisite conditions to proper accelerations to reach, e. g., $\sim 10^{20}$ eV cosmic rays \cite{takahara,rachen,celotti}.
In this manner, the presented results could possibly contribute to constrain magnetic and dynamic properties of AGNs to better understand the requisite conditions to proper accelerate 
the highest energy cosmic rays.

\begin{acknowledgments}

The authors are very grateful to the researchers of DEE-UFPR. This work was funded by CNPq under grant 458896/2013-6. 

\end{acknowledgments}

\begin{table}[p]
\centering
\caption{Comparison between cosmic ray luminosity $L_{pr}^{UL}$ (protons) from the method derived from eq. (\ref{eq:CRUL}) and the  
theoretically calculated range $L_{CRmin}^{Theory}-L_{CRmax}^{Theory}$
of cosmic ray luminosities to $a=0.45$ and $a=0.7$, from many sources. For comparison, it is also calculated the upper limit of 
the bolometric luminosity (the Eddington luminosity $L_{\mathrm{Edd}}$) for each case below, in erg s$^{-1}$: $L_{\mathrm{Edd}}[\mathrm{NGC}985]=5.02\times10^{48}$, 
$L_{\mathrm{Edd}}[\mathrm{NGC}1142]=6.31\times10^{48}$, $L_{\mathrm{Edd}}[2\mathrm{MASXJ}07]=4.68\times10^{48}$, $L_{\mathrm{Edd}}[\mathrm{CGCG}420]=6.83\times10^{48}$, 
$L_{\mathrm{Edd}}[\mathrm{MCG}-01]=1.82\times10^{48}$, $L_{\mathrm{Edd}}[2\mathrm{MASXJ}11]=1.26\times10^{48}$, $L_{\mathrm{Edd}}[\mathrm{LEDA}]=4.92\times10^{48}$, 
$L_{\mathrm{Edd}}[\mathrm{NGC5995}]=9.78\times10^{47}$, $L_{\mathrm{Edd}}[\mathrm{Mrk520}]=3.17\times10^{48}$. 
The mass source comes from
  \cite{lisa,michael}. Redshift and other properties are taken from \cite{oh}.}
\label{tab:1}
\begin{tabular}{|c|c|c|c|c c|c|c|}
\hline \hline & & & & & & $a=0.45$ & $a=0.7$\\
\textbf{Source name} & \textbf{$z_s$} & \textbf{\hbox{log}$M_{\odot}$}  & $\mathbf{L_{pr}^{UL}}$ (Proton)  & \multicolumn{2}{|c|}{$\mathbf{L_{B}}$[erg s$^{-1} \times 10^{45}$]}  & $\mathbf{L_{CRmin}^{Theory}-L_{CRmax}^{Theory}}$ & $\mathbf{L_{CRmin}^{Theory}-L_{CRmax}^{Theory}}$ \\
			    &                             &
                            &[erg s$^{-1} \times 10^{45}$]& $a = 0.45$ &$a = 0.7$ & [erg s$^{-1} \times 10^{45}$] & [erg s$^{-1} \times 10^{45}$]\\ \hline
    {\small NGC 985}       & 0.04353 &10.6      &         1.03      &5.14 &15.54& 0.21 - 2.57& 0.77 - 2.79\\ \hline
  {\small NGC 1142}      & 0.02916 & 10.7     &         0.49      &6.47  &19.57& 0.26 - 3.23& 0.97 - 3.52\\ \hline
  {\small 2MASX J07595347+2323241}     &0.03064  & 10.57 &        1.01      &4.80&14.51& 0.19 - 2.40& 0.72 - 2.61\\ \hline
   {\small CGCG 420-015}  & 0.02995 & 10.63      &         0.95      & 5.51&16.66 & 0.22 - 2.76& 0.83 - 2.99\\ \hline
    {\small MCG-01-24-012} &0.02136  & 10.16       &         0.65      &  1.86 &5.64 & 0.07 - 0.93& 0.28 - 1.02\\ \hline
  {\small 2MASX J11454045-1827149}  &0.03616  & 10.0 &       1.30      &1.29 &3.90 &  0.07 - 0.64 & 0.19 - 0.70   \\ \hline
  {\small LEDA 170194} & 0.04024 &  10.59      &         1.48      & 5.02 &15.19 & 0.21 - 2.51& 0.75 - 2.73\\ \hline
   {\small NGC 5995}   & 0.02834  &  9.89      &         0.90      & 1.00 &3.03 & 0.04 - 0.51& 0.15 - 0.54\\ \hline
   {\small Mrk 520}   & 0.02772  &  10.4      &         0.98      & 3.24 &9.81 & 0.13 - 1.62& 0.49 - 1.77\\ \hline
 \end{tabular}
\end{table}
\end{document}